\documentclass[prl,amsmath,amssymb, twocolumn, showpacs]{revtex4}

\usepackage{dcolumn}

\usepackage{bm}

\usepackage[dvips]{graphicx}

\usepackage{epsfig}

\begin{document}

\title{Four-level and two-qubit systems, sub-algebras, and unitary integration}

\author{A.\ R.\  P. Rau$^{1,*}$, G.\ Selvaraj$^1$ and D.\ Uskov$^{1,2}$}
\affiliation{$^1$Department of Physics and Astronomy, Louisiana State University,
Baton Rouge, Louisiana 70803-4001, USA.\\
$^2$ Department of Optics, P.N.Lebedev Physicsl Institute, 119991 Moscow, Russia.}


\begin{abstract}

Four-level systems in quantum optics, and for representing two qubits in quantum computing, 
are difficult to solve for general time-dependent 
Hamiltonians. A systematic procedure is presented which combines 
analytical handling of the algebraic operator aspects with simple 
solutions of classical, first-order differential equations. In 
particular, by exploiting $su(2) \oplus su(2)$ and $su(2) \oplus 
su(2) \oplus u(1)$ sub-algebras of the full $SU(4)$ dynamical group of the 
system, the non-trivial part of the final calculation is reduced to a 
single Riccati (first order, quadratically nonlinear) equation, 
itself simply solved. Examples are provided of two-qubit problems from the 
recent literature, including implementation of two-qubit gates with Josephson junctions.

\end{abstract}


\maketitle

Four-level systems are often of interest in quantum optics and in 
multiphoton processes. As examples of recent literature, see, for 
instance, \cite{ref1,ref2}. Added interest is provided by recent 
applications for quantum information and computing where two qubits 
constitute a four-level system. Indeed, two qubits and associated 
logic gates have been practically realized through Josephson 
junctions \cite{ref3}, the coupling Hamiltonian viewed as a $4 \times 
4$ matrix.

In recent papers, we have developed a ``unitary integration" 
technique for solving time-dependent operator equations such as the 
Schr\"{o}dinger or Liouville equation for the state or density matrix 
of a quantum system \cite{ref4}. The method also extends to more 
general master equations when dissipation and decoherence are present 
\cite{ref5}. The evolution operator is written in the form of a 
product of exponentials, each exponent itself a product of a 
classical function of time and one quantum operator. When these 
operators form a closed algebra under commutations, first-order 
time-dependent equations can be derived for the introduced classical 
functions, and their resulting solutions allow the construction of a 
complete solution of the quantal problem. The method works best, of 
course, when the number of operators in the algebra is small. For one 
qubit, where the most general Hamiltonian involves only four 
operators, the three Pauli matrices and the unit matrix, the $su(2)$  
algebra has just three elements. Upon including dissipation and 
decoherence, a $3 \times 3$ matrix for the elements of the density 
matrix introduces, at its most general, eight elements of an $su(3)$ 
algebra \cite{ref5}. For special situations when an $su(2)$ sub-algebra 
suffices, such decoherence problems can also be treated through just 
three classical functions whose defining equations further reduce to 
just one non-trivial Riccati (first-order, quadratically nonlinear) 
equation (or, equivalently, to a linear second-order differential 
equation).

The aim of this paper is to examine two-qubit or four-level problems 
analogously. The complete description in terms of the 15 operators of 
the symmetry group $SU(4)$ (together with the unit operator) requires 
15 exponential factors in the evolution operator which is cumbersome. 
However, judicious choices of sub-algebras reduce the problems 
considerably. These may be of interest since they describe several of 
the two-qubit logic gates and four-level systems in the recent 
literature. Specifically, we present here the cases pertaining to two 
sub-algebras, $su(2) \oplus su(2)$ and $su(2) \oplus su(2) \oplus 
u(1)$. In both instances, the same single Riccati equation for a 
single function is all that remains to be solved (through, for example, a popular Mathematica package \cite{ref6}) to provide complete descriptions of such 
systems. Other sub-algebras, namely $so(5)$ with ten operators and $su(3)$ 
with eight, will be the subject of future work.

\section{Algebra of operators for a four-state system}

\begin{table*}
\begin{center}
\begin{tabular}{|c||c|c|c|c|c|c|c|c|c|c|c|c|c|c|c|c|}

\hline
$O_X $&$O_2 $&$O_3  $&$O_4  $&$O_5   $&$O_6    $&$O_7   $&$O_8  $&$O_9   $&$O_{10}  $&$O_{11}  $&$O_{12}   $&$O_{13}  $&$O_{14}    $&$O_{15}   $&$ O_{16}  $\\ \hline \hline
$O_2 $&$0   $&$0    $&$0    $&$iO_6  $&$-iO_5  $&$iO_8 $&$-iO_7$&$0     $&$0     $&$0     $&$0      $&$iO_{16} $&$-iO_{15}$&$iO_{14}  $&$-iO_{13}  $\\
\hline
$O_3 $&$0   $&$0    $&$ 0    $&$0     $&$0      $&$0     $&$ 0 $&$iO_{10} $&$-iO_9 $&$iO_{12} $&$-iO_{11} $&$iO_{15} $&$-iO_{16}  $&$-iO_{13} $&$ iO_{14} $\\ \hline
$O_4 $&$0   $&$0    $&$0    $&$iO_8  $&$-iO_7  $&$\frac{i}{4}O_6$&$\frac{-i}{4}O_5$&$iO_{12} $&$-iO_{11}$&$\frac{i}{4}O_{10}$&$-\frac{i}{4}O_9$ &$0$&$0$&$0$&$0$\\
\hline
$O_5 $&$-iO_6$&$ 0 $&$ -iO_8 $&$ 0 $&$ iO_2 $&$ 0$&$ iO_4 $&$ 0 $&$ 0 $&$-iO_{16}$&$-iO_{14}$&$ 0 $&$iO_{12}$&$0$&$iO_{11}$\\
\hline
$O_6 $&$iO_5 $&$ 0 $&$iO_7$&$-iO_2$&$0$&$ 0 $&$-iO_4$&$0$&$ 0 $&$ 0 $&$iO_{15}$&$-iO_{11}$&$0$&$-iO_{12}$&$ 0 $\\
\hline
$O_7 $&$-iO_8 $&$0$&$\frac{-i}{4}O_6$&$0$&$iO_4$&$0$&$\frac{i}{4}O_2$&$iO_{15}$&$-iO_{13}$&$0$&$0$&$\frac{i}{4}O_{10}$&$0$&$\frac{-i}{4}O_9$&$0$\\ \hline
$O_8 $&$iO_7 $&$ 0 $&$\frac{i}{4}O_5$&$-iO_4$&$ 0 $&$\frac{-i}{4}O_2$&$0$&$iO_{14}$&$-iO_{16}$&$0$&$0$&$0$&$\frac{-i}{4}O_9$&$0$&$\frac{i}{4}O_{10}$\\
\hline
$O_9 $&$0   $&$-iO_{10}$&$-iO_{12}$&$0$&$0$&$-iO_{15}$&$-iO_{14}$&$0$&$iO_3$&$0$&$iO_4$&$0$&$iO_8$&$iO_7$&$0$\\
\hline
$O_{10} $&$0   $&$iO_9$&$iO_{11}$&$0$&$0$&$iO_{13}$&$iO_{16}$&$-iO_3$&$0$&$-i0_4$&$0$&$-iO_7$&$0$&$0$&$-iO_8$\\
\hline
$O_{11}$&$ 0 $&$-iO_{12}$&$\frac{-i}{4}O_{10}$&$iO_{16}$&$-iO_{13}$&$0$&$0$&$0$&$iO_4$&$0$&$\frac{i}{4}O_3$&$\frac{i}{4}O_6$&$0$&$0$&$\frac{-i}{4}O_5$\\
\hline
$O_{12}$&$0$&$iO_{11}$&$\frac{i}{4}O_9$&$iO_{14}$&$-iO_{15}$&$0$&$0$&$-iO_4$&$0$&$\frac{-i}{4}O_3$&$0$&$0$&$\frac{-i}{4}O_5$&$\frac{i}{4}O_6$&$0$\\
\hline
$O_{13}$&$-iO_{16}$&$-iO_{15}$&$0$&$0$&$iO_{11}$&$\frac{-i}{4}O_{10}$&$0$&$0$&$iO_7$&$\frac{-i}{4}O_6$&$0$&$0$&$0$&$\frac{i}{4}O_3$&$\frac{i}{4}O_2$\\
\hline
$O_{14}$&$iO_{15}$&$iO_{16}$&$0$&$-iO_{12}$&$0$&$0$&$\frac{i}{4}O_9$&$-iO_8$&$0$&$0$&$\frac{i}{4}O_5$&$0$&$0$&$\frac
{-i}{4}O_2$&$\frac{-i}{4}O_3$\\
\hline
$O_{15}$&$-iO_{14}$&$iO_{13}$&$0$&$0$&$i0_{12}$&$\frac{i}{4}O_9$&$0$&$-iO_7$&$0$&$0$&$\frac{-i}{4}O_6$&$\frac{-i}{4}O_3$&$\frac{i}{4}O_2$&$0$&$0$\\
\hline
$O_{16}$&$iO_{13}$&$-iO_{14}$&$0$&$-iO_{11}$&$0$&$0$&$\frac{-i}{4}O_{10}$&$0$&$iO_8$&$\frac{i}{4}O_5$&$0$&$\frac{-i}{4}O_2$&$\frac{i}{4}O_3$&$0$&$0$\\
\hline
\end{tabular}
\end{center}
\caption{Table of commutators. With operators $O_i$ in the first column and $O_j$ in the top row, each entry provides the commutator $[O_i,O_j]$.}
\end{table*}

\subsection{$SU(2) \times SU(2)$ sub-groups}

The sixteen linearly independent operators of a four-state system can 
be chosen in a variety of matrix representations. One choice, 
familiar in nuclear and particle physics applications \cite{ref7}, 
are the 15 $\lambda_i$ matrices of $su(4)$. This set is suitable for 
examining $su(3)$ and $su(2)$ sub-algebras involving 8 and 3 operators, 
respectively, but for our interest in this paper, an alternative 
representation proves more convenient. This starts from two qubits 
and employs their individual Pauli matrices, 6 in number, along with 
tensor products, 9 in number: $\frac{1}{2} \vec {\sigma}, \frac{1}{2} \vec 
{\tau}, \frac{1}{2} \vec{\sigma} \otimes \frac{1}{2} \vec{\tau}$. 
Together with the unit $4 \times 4$ matrix, these 16 matrices called 
$O_i$ are explicitly tabulated in an earlier paper \cite{ref8}. (A 
slightly different arrangement of the same operators, called $X_i$, 
with much useful group-theoretic discussion, is in \cite{ref9}). They 
have the advantage for the unitary integration procedure (to be 
described below) that the square of each is, to within a factor, the 
unit matrix. Exponentiation of these operators is thereby rendered 
simple, involving just two terms.

Table I gives the commutators 
$[O_i, O_j]$ and is useful for picking out sub-algebras. For example, $[O_2, O_5]=iO_6$. The four 
states in this representation may be described either in the 
four-state language as the column vectors $(1,0,0,0), (0,1,0,0), 
(0,0,1,0), (0,0,0,1)$, or in spin notation as $|\uparrow \uparrow 
\rangle, |\uparrow \downarrow \rangle, |\downarrow \uparrow \rangle, 
|\downarrow \downarrow \rangle $, respectively.

There is a large set of six-dimensional sub-algebras formed out 
of two triplets obeying $su(2)$ or, equivalently, $so(3)$ angular momentum commutation 
relations. One such set, $(O_2, O_5, O_6; O_3, O_9, O_{10})$, is the trivial $su(2)\oplus su(2)$ subalgebra of local operations on two qubits $\vec{\sigma}$ and $\vec{\tau}$. There are many equivalent sets generated by a Similarity Transformation (ST) acting on the elements of a subalgebra by right-left multiplication by an arbitrary element of the full $SU(4)$ group. Let $W$ be a unitary $4 \times 4$ matrix, and $x$ an $su(4)$ algebra element. Then, ST acts as 
\[
x' = WxW^{ - 1}  \equiv WxW^\dag .  
\]
\noindent  In this section we will use only a discrete subgroup of all possible ST transformations, which have the property that the set of 15 operators $O_i$ is transformed into the same set $O_j$, i.e. the action of ST is equivalent to a permutation among the $O_i$. In this case the $su(2)\oplus su(2)$ subalgebras, obtained by ST transformation of $\vec{\sigma}, \vec{\tau}$ can be identified by direct analysis of subgroup structures using Table 1.

Subgroups different from the trivial $\vec{\sigma},\vec{\tau}$ one  can be generated only by two-qubit entangling operations, such as transformation to the Bell basis (or magic Bell basis \cite{ref10}). One can easily verify that the set $(O_9, O_8, 
O_{14}; O_6, O_{11}, O_{13})$ is such a subalgebra. That is, the operator sets $(\tau_x, \sigma_y \tau_z, \sigma_y \tau_y)$ and  $(\sigma_y, \sigma_z \tau_x, \sigma_x \tau_x)$  are two mutually commuting 
triplets with $su(2)$ commutation relations. The ST 
transformation with the unitary matrix
\begin{equation}
\frac{1}{{\sqrt 2 }}\,\,\left( {\begin{array}{*{20}c}
   1 & 0 & 0 & 1  \\
   0 & 1 & 1 & 0  \\
   0 & 1 & { - 1} & 0  \\
   1 & 0 & 0 & { - 1}  \\
\end{array}} \right)
\label{eqn1}
\end{equation}
passes from the $\vec{\sigma}$ set to $(\tau_x, \sigma_y 
\tau_z, \sigma_y \tau_y)$ and from $\vec{\tau}$ to $(\sigma_y, \sigma_z \tau_x, \sigma_x 
\tau_x)$. The basis corresponding to Eq.~(\ref{eqn1}) is the Bell basis, 
$(1,0,0,1)=|\uparrow \uparrow \rangle + | \downarrow \downarrow 
\rangle , (0,1,1,0)= |\uparrow \downarrow ~\rangle + |\downarrow 
\uparrow \rangle , (0,1,-1,0)= |\uparrow \downarrow \rangle - 
|\downarrow \uparrow \rangle , (1,0,0,-1)=|\uparrow \uparrow \rangle 
- | \downarrow \downarrow \rangle $, which are simultaneous eigenstates of $\sigma_x \tau_x$ 
and $\sigma_y \tau_y$ \cite{ref10}.

All such subalgebras are maximal $su(2)\oplus su(2)$ sub-algebras of  
$su(4)$, in the sense that, if any operator from Table I 
is added to the six already chosen, the algebra closes (as a Lie algebra) only in the full $su(4)$ algebra.
Another perspective on the above Bell basis is provided by 
considering a four-level system with nearest neighbor couplings as in 
the Hamiltonian,

\begin{equation}
H = \left( {\begin{array}{*{20}c}
   0 & \alpha  & 0 & 0  \\
   \alpha  & 0 & \beta  & 0  \\
   0 & \beta  & 0 & \gamma   \\
   0 & 0 & \gamma  & 0  \\
\end{array}} \right)
\label{eqn2}
\end{equation}
Such a matrix may be viewed as built up of three symmetric $4 \times 4$ matrices $m_{jk}$,  with unit entries in row (column) $j$ and column(row) $k$ and zero everywhere else,  so 
that $H=\alpha m_{12}+\beta m_{23} + \gamma m_{34}$.  Starting with these three real, symmetric matrices, we define two non-zero commutators between them, $m_{13}=-i[m_{12},m_{23}], m_{24}=-i[m_{23},m_{34}]$ to give two imaginary, anti-symmetric matrices, and a final real symmetric $m_{14}=-i[m_{13},m_{34}]$ to complete a full 6-dimensional $so(4)$ algebra. Using the well-known relation between $SO(4)$ and $SU(2) \times SU(2)$ groups, this sub-algebra decomposes into two sets of mutually commuting triplets,
$m_{12}+m_{34}, m_{13}-m_{24}, m_{12}-m_{34}$ and $m_{12}-m_{34}, m_{13}+m_{24}, m_{23}-m_{14}$. These linear  combinations correspond to the subalgebra $(O_9, O_8, O_{14}; O_6, O_{11}, O_{13})$ of the full $su(4)$ algebra. Such a construction generalizes to $n$-level systems, $n>4$, with nearest neighbor couplings. The matrices $m_{jk}$ defined as

\[
(m_{jk})_{pq} = (-i)^{|j-k|-1} [\delta_{jp} \delta_{kq} +(-1)^{k-j-1} \delta_{jq} \delta_{kp}],
\]
provide $n(n-1)/2$ generators of a $so(n)$ algebra. These form a sub-algebra of the $n^2 -1$-dimensional $su(n)$ algebra although, for $n>4$, the $so(n)$ no longer reduces to a product of $su(2)$ algebras.

\subsection{$SU(2) \times SU(2) \times U(1)$ sub-groups}
A class of maximal seven-dimensional sub-algebras of $su(4)$, different from the one considered  in the previous subsection, can be constructed by taking linear combinations of $O_n$ generators.  Inspection of Table I shows that  the first seven elements form a  diagonal sub-block closed under commutation, isomorphic to a $so(4)$ algebra. With $O_3$ 
commuting with all the rest of this sub-block, the other six divide into two mutually 
commuting $su(2)$ triplets upon forming appropriate linear combinations which pass from a standard $4 \times 4$ antisymmetric real-matrix representation of $so(4)$ to $su(2)\oplus su(2)$ matrix ($4 \times 4$) form.
Another such example, previously considered by one of us \cite{ref8}, is 
provided by $(O_2, O_3, O_{13}, O_{14}, O_{15}, O_{16})$, with $O_4$ 
commuting with all of them. The linear combinations, defined in 
\cite{ref8} as ``pseudo-spins",  obey 
$su(2)$ commutation relations while not being isomorphic to Pauli spinors. While the Pauli matrices constitute a matrix algebra with the unit operator as the identity element, the pseudo-spin matrix algebra has as its identity element the operator $\frac{1}{2}\left( {\hat I \pm \sigma _z \tau _z} \right)$. One implication is that this particular type of $su(2)\oplus su(2)$ algebra is not related by a ST operation to the $su(2)\oplus su(2)$ algebras considered in the previous section.    
These $su(2)\oplus su(2)\oplus u(1)$ sets arise, for example, in 
the physical context of nuclear 
magnetic resonance \cite{ref8} with scalar coupling and heteronuclear two-spin 
coherences. For applications in methods such as unitary integration 
which require closure under commutation, such pseudo-spins work 
equally well as we will see below. There are many such sub-algebras which can be obtained by the discrete subgroup of ST operations as discussed in the previous section. For example, one can simply relabel  $(x,y,z)$. Indeed, 
each row (or column) of Table I has seven zero entries. The 
corresponding seven operators constitute such an algebra for a total 
of 15 sets \cite{ref11}.

The recent demonstration of two qubit gates formed out of 
Josephson junctions \cite{ref3} considers the Hamiltonian,

\begin{equation}
H = \left(
\begin{array}{cccc}
E_{00} & -\frac{1}{2} E_{J1} & -\frac{1}{2} E_{J2} & 0 \\
-\frac{1}{2} E_{J1} & E_{10} & 0 & -\frac{1}{2} E_{J2} \\
-\frac{1}{2} E_{J2} & 0 & E_{10} & -\frac{1}{2} E_{J1} \\
0 & -\frac{1}{2} E_{J2} & -\frac{1}{2} E_{J1} & E_{00}
\end{array}
\right).
\label{eqn4}
\end{equation}
Expressing in terms of the $O_i$ matrices in Table I, this 
Hamiltonian is a linear combination of $(O_1, O_4, O_5, O_9)$. 
Together with $(O_8, O_{12}, O_{14})$, they form a closed 
sub-algebra, and the operator $O_{13}$ commutes with all of them. For 
applications below in Section 2, we record the Hamiltonian in terms 
of the Pauli spinors,

\begin{equation}
H = \frac{1}{2} (E_{00} +E_{10}) - \frac{1}{2} (E_{J2} \sigma_x 
+E_{J1} \tau_x) +\frac{1}{2} (E_{00}-E_{10}) \sigma_z \tau_z,
\label{eqn5}
\end{equation}
and

\begin{eqnarray}
H & = & \frac{1}{2} (E_{00}+E_{10}) \mathcal{I} +\frac{1}{2} 
(\omega_{+} S_z +\omega_{-} s_z) \nonumber \\
  & + & \!\! \frac{1}{4} (E_{00}-E_{10})(s_{+}\! +s_{-}\! -S_{+}\! -S_{-}\!),
\label{eqn6}
\end{eqnarray}
where we have defined $\omega_{\pm} =-E_{J2} \mp E_{J1}, S_z 
=\frac{1}{2} (\sigma_x + \tau_x), s_z =\frac{1}{2} (\sigma_x - 
\tau_x)$, and

\begin{equation}
S_{\pm} =\frac{1}{2} (\sigma_y \pm i\sigma_z)(\tau_y \pm i\tau_z), 
\,\, s_{\pm}=\frac{1}{2} (\sigma_y \pm i\sigma_z)(\tau_y \mp i\tau_z).
\label{eqn7}
\end{equation}
The sets $(S_z, S_{\pm})$ and $(s_z, s_{\pm})$ mutually commute and 
each obeys the commutation relations of an $su(2)$ algebra but, as in 
\cite{ref8}, they are not completely isomorphic to the $4 \times 4$ matrix $\sigma_i$ or $\tau_i$ algebras, as discussed above. In 
particular, as is easily seen, the squares of $S_z$ and $s_z$ are not 
the unit operator but involve $\sigma_x \tau_x$ as well. While 
sharing the same commutation relations as the similar pseudo-spins 
defined in \cite{ref8}, the sets here differ as $4 \times 4$ 
matrices, being in the ``$x$-representation" instead of the 
counterpart ``$z$-representation" in \cite{ref8}.

\section{Unitary integration of the evolution operator}

Time-dependent quantum problems can be conveniently handled through 
the unitary integration procedure of \cite{ref4}. Solutions of the 
Schr\"{o}dinger equation for the quantum state are obtained through 
the evolution operator $U(t)$ defined by

\begin{equation}
i\dot{U}(t)=H(t)U(t),\quad U(0)=\mathcal{I},  \label{eqn8}
\end{equation}
\noindent where an over-dot denotes differentiation with respect to time.
For more general treatment that includes mixed states, master 
equations such as the Liuoville-von Neumann-Lindblad equation 
\cite{ref12},

\begin{eqnarray}
i\dot{\rho} & = & [H,\rho ] +
\frac{1}{2}i\!\sum_{k}\left( [L_k\rho,L_k^{\dagger }]+
[L_k,\rho L_k^{\dagger }]\right)  \nonumber \\
  & \!\!= \!\!& [H,\rho ] \!- \!\frac{1}{2}i\!\sum_{k}\left( 
L_k^{\dagger }L_k\rho +\rho
L_k^{\dagger }L_k-2L_k\rho L_k^{\dagger }\right)\!, \label{eqn9}
\end{eqnarray}
when first cast in terms of the individual density matrix elements, 
take a form similar to Eq.~(\ref{eqn8}), namely

\begin{equation}
i\dot{\eta}(t)=\mathcal{L} \eta (t),
\label{eqn10}
\end{equation}
where $\eta$ are appropriate linear combinations of matrix elements of the density matrix $\rho_{ij}$ \cite{ref5}, such as, for example, the ones obtained by a decomposition of $\rho_{ij}$ in the set of generators of the corresponding $SU(n)$ group \cite{ref13}. We have
 
 \begin{eqnarray}
 \eta & = & [ \rho_{11}-\rho_{22}, (\rho_{11} +\rho_{22} -2\rho_{33})/\sqrt{3}, \nonumber \\ 
 & & (\rho_{11} +\rho_{22} +\rho_{33} -3\rho_{44})/\sqrt{6}, \rho_{12}+\rho_{21}, \nonumber \\ 
 & & i(\rho_{12}-\rho_{21}),  \rho_{13}+\rho_{31}, i(\rho_{13}-\rho_{31}), \rho_{14}+\rho_{41},  \nonumber \\
 & & i(\rho_{14}-\rho_{41}), \rho_{23}+\rho_{32}, i(\rho_{23}-\rho_{32}), \rho_{24}+\rho_{42},  \nonumber \\
 & & i(\rho_{24}-\rho_{42}), \rho_{34}+\rho_{43}, i(\rho_{34}-\rho_{43})].
 \label{eqn11}
 \end{eqnarray} 

\begin{figure}[h]
\includegraphics[width=3in]{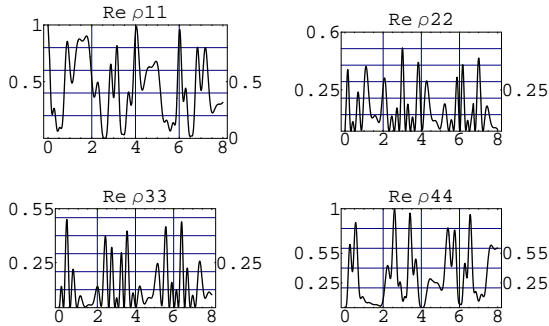}
\caption{Time evolution of the diagonal elements of the density matrix of a $n=4$ two-qubit Josephson junction system. The horizontal axis is ($\omega t/2\pi $), parameters as given in the text with $\omega = 1$ GHz, $\delta=0$, and the initial configuration $\rho (0)=\delta_{i1} \delta_{j1}$.}
\end{figure}

\begin{figure}[h]
\includegraphics[width=3in]{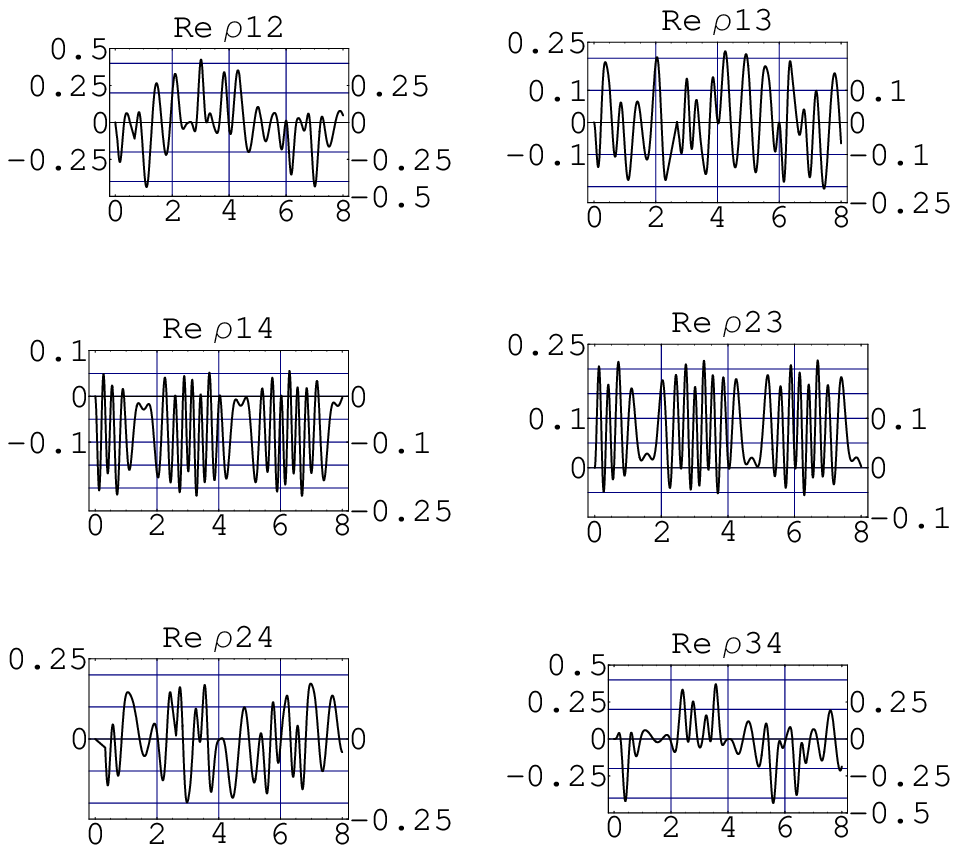}
\caption{As in Fig.1 but for the real part of the off-diagonal elements of the density matrix.}
\end{figure}

\begin{figure}[h]
\includegraphics[width=3in]{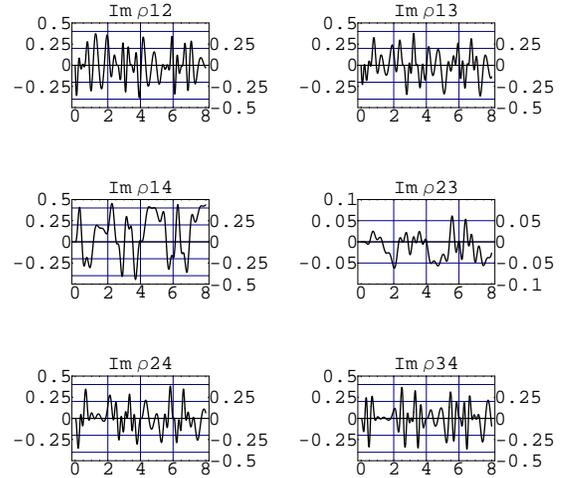}
\caption{As in Fig.1 but for the imaginary part of the off-diagonal elements of the density matrix.}
\end{figure}

In either case of Eq.~(\ref{eqn8}) or Eq.~(\ref{eqn10}), unitary 
integration writes the solution as a product of exponentials, 
$\prod_{j}\exp [-i\mu _j(t)A_j]$, with time-dependent, classical 
functions $\mu _j(t)$, and $A_j$ the elements of a complete algebra 
\cite{ref4}. Upon substituting this form into Eq.~(\ref{eqn8}) or 
Eq.~(\ref{eqn10}), a consistent set of defining equations for the 
$\mu_j$ follow so long as each element of the algebra is included in 
the above product. As shown in \cite{ref4,ref5}, so long as one has 
an $su(2)$ algebra, solving this set defining $\mu_j$ reduces to just 
one non-trivial step, the solution of a single Riccati equation. This 
remains true if more than one $su(2)$ is involved so long as they are 
mutually commuting, each set involving one Riccati equation. Note 
also that since only commutators of $A_j$ are involved in the 
derivation, the results apply also to pseudo-spins as in the previous 
section, or for other algebras such as $so(3)$ or $su(1,1)$ differing 
only in the sign or value of the structure coefficients which do not 
change the form of the defining equations for the $\mu$.

\begin{figure}[h]
\includegraphics[width=3in]{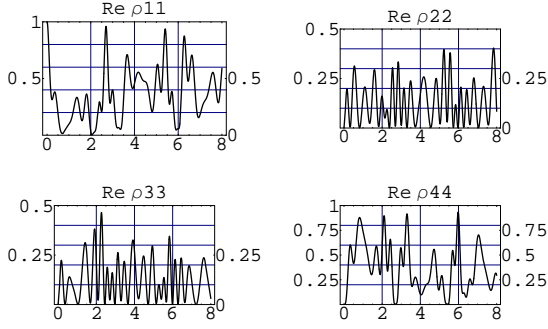}
\caption{Time evolution of the diagonal elements of the density matrix of a $n=4$ two-qubit Josephson junction system. The horizontal axis is ($\omega t/2\pi $), parameters as given in the text with $\omega = 1$ GHz, $\delta=\pi /4$, and the initial configuration $\rho (0)=\delta_{i1} \delta_{j1}$. Contrast with Fig. 1 to see differences due to a mutual phase between the driving fields.}
\end{figure}

\begin{figure}[h]
\includegraphics[width=3in]{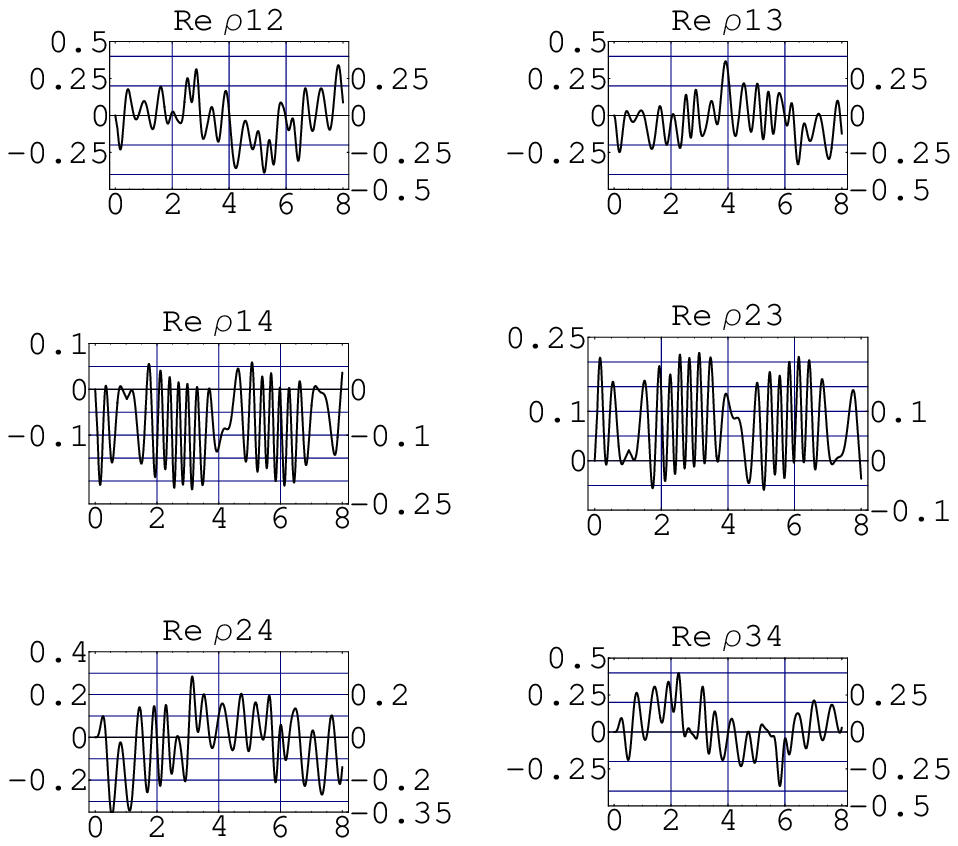}
\caption{As in Fig. 4 but for the real part of the off-diagonal elements of the density matrix.}
\end{figure}

\begin{figure}[h]
\includegraphics[width=3in]{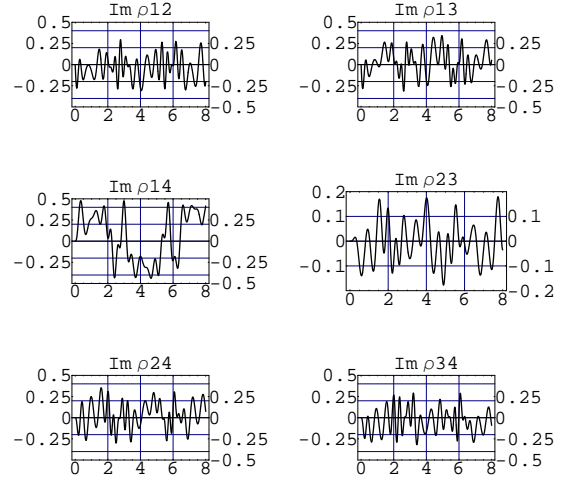}
\caption{As in Fig. 4 but for the imaginary part of the off-diagonal elements of the density matrix.}
\end{figure}

It is this feature that we now exploit in applying to four-level or 
two qubit problems that fall into the two categories of the $su(2) 
\oplus su(2)$ and $su(2)\oplus su(2)\oplus u(1)$ sub-algebras 
discussed in Section 2. An example of the former is a recent discussion \cite{ref14} of geometric phases with two qubits. More general time dependences than considered in that reference are amenable to our procedure presented below. For the coupled pair of Josephson junctions 
of \cite{ref3} with the Hamiltonian in Eq.~(\ref{eqn6}), we have from 
our previous derivation in \cite{ref5} the evolution operator

\begin{eqnarray}
U(t) &\!\!\!\! =\!\!\!\! &\exp(-i\Omega_0(t)) \exp[-\frac{1}{2} 
i\nu_{+}(t)S_{+}] \exp[-\frac{1}{2} i\nu_{-}(t)S_{-}] \nonumber \\
        & \times & \exp[-\frac{1}{2} i\nu_3(t)S_z] \exp[-\frac{1}{2} 
i\mu_{+}(t)s_{+}]  \nonumber \\
         & \times & \exp[-\frac{1}{2} i\mu_{-}(t)s_{-}] 
\exp[-\frac{1}{2} i\mu_3(t)s_z].
\label{eqn12}
\end{eqnarray}
The sets $\mu$ and $\nu$ are for the two $SU(2)$'s involved in problems 
such as those in \cite{ref3}. An additional term in $\sigma_x \tau_x$ 
in Eq.~(\ref{eqn6}) could also be accommodated through an additional 
exponential factor in Eq.~(\ref{eqn12}) and would correspond to an 
U(1).
These classical functions in the exponents are obtained through the 
set of equations \cite{ref5},

\begin{eqnarray}
\dot{\Omega_0} & = & (E_{00} + E_{10})/2,  \nonumber \\
\dot{\mu_{+}} - {\mu_{+}}^2 k_{-} +i\omega_{-} \mu_{+} & = & k_{+}, 
\nonumber \\
\dot{\mu_{-}} -i\mu_{-}\dot{\mu_3} =k_{-},  &  & \dot{\mu_3} 
-2ik_{-}\mu_{+} =\omega_{-},
\label{eqn13}
\end{eqnarray}
with a similar set for the functions $\nu$ which have $K_{\pm}$ 
instead of $k_{\pm}$. These parameters for the $H$ in 
Eq.~(\ref{eqn6}) take the values $K_{\pm} =-k_{\pm} 
=(E_{00}-E_{10})/2$. For applications with $k$ constant  as in \cite{ref3},  
and $\omega_{\pm}$ functions of time, the first, and 
only non-trivial, equation of the set in Eq.~(\ref{eqn13}) can also 
be recast as \cite{ref4},

\begin{equation}
\dot{\theta}(t) = k-\frac{1}{2}\omega_{-}(t)\sin 2\theta(t),
\label{eqn14}
\end{equation}
with $\mu_{+} \equiv \tan (\theta +kt)$, or

\begin{equation}
\ddot{\gamma}(t) +i\omega_{-}(t) \dot{\gamma}(t) +k^2 \gamma(t)=0,
\label{eqn15}
\end{equation}
with $\mu_{+} \equiv -(1/k)(d/dt)(\ln \gamma)$.
In any of these alternative forms, $\mu_{+}$ (and, similarly, 
$\nu_{+}$) can be solved for a given $\omega_{\pm}$, following which 
the rest of the set in Eq.~(\ref{eqn13}) yields to simple integration.

In Figs. 1-3 , we present results for the two-qubit Josephson junction system of \cite{ref3}. The parameters of that study were $E_{00}-E_{10}=7.85$ GHz, $E_{J1}=13.4$ and $E_{J2}=9.1$ GHz. All these parameters were held constant in that study but, with an eye to future applications where these energies may vary with time, we consider such applications of our general time-dependent formalism. Specifically, were the voltages to be harmonically modulated, with $E_{J1}=13.4 \cos (\omega t+\delta)$ and $E_{J2}=9.1 \cos (\omega t)$, solution of  Eq.~(\ref{eqn13}) leads to diagonal and off-diagonal components of the density matrix shown in Figs. 1-3. Starting with an initial pure state $(1,0,0,0)$, the density matrix, $\rho(t) = U \rho(0) U^{\dagger}$ is evaluated and exhibits complicated oscillatory population of the four levels (diagonal density matrix elements in Fig. 1 which, of course, always sum to unity for all $t$) as also coherences (off-diagonal elements in Figs. 2 and 3). Figs. 4-6 show the effect of a phase difference between the two driving fields, leading to interesting differences when compared to Figs. 1-3 where the driving fields are in phase.

\begin{figure}[h]
\includegraphics[width=3in]{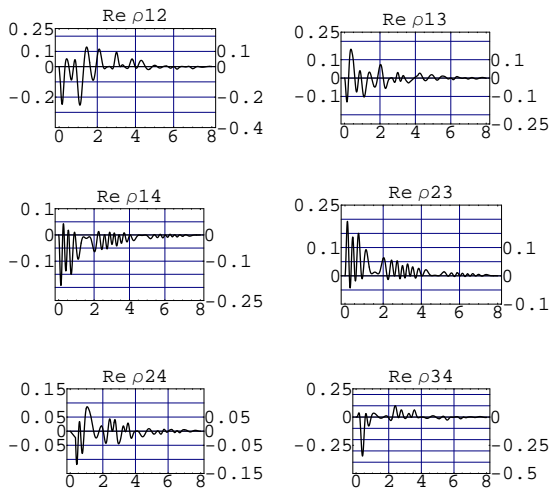}
\caption{As in Fig. 1 but with a damping constant $\Gamma = 0.5$ GHz. Note the differences from Fig. 1 as the oscillations damp out to give asymptotic values of 1/4.}
\end{figure} 

\begin{figure}[h]
\includegraphics[width=3in]{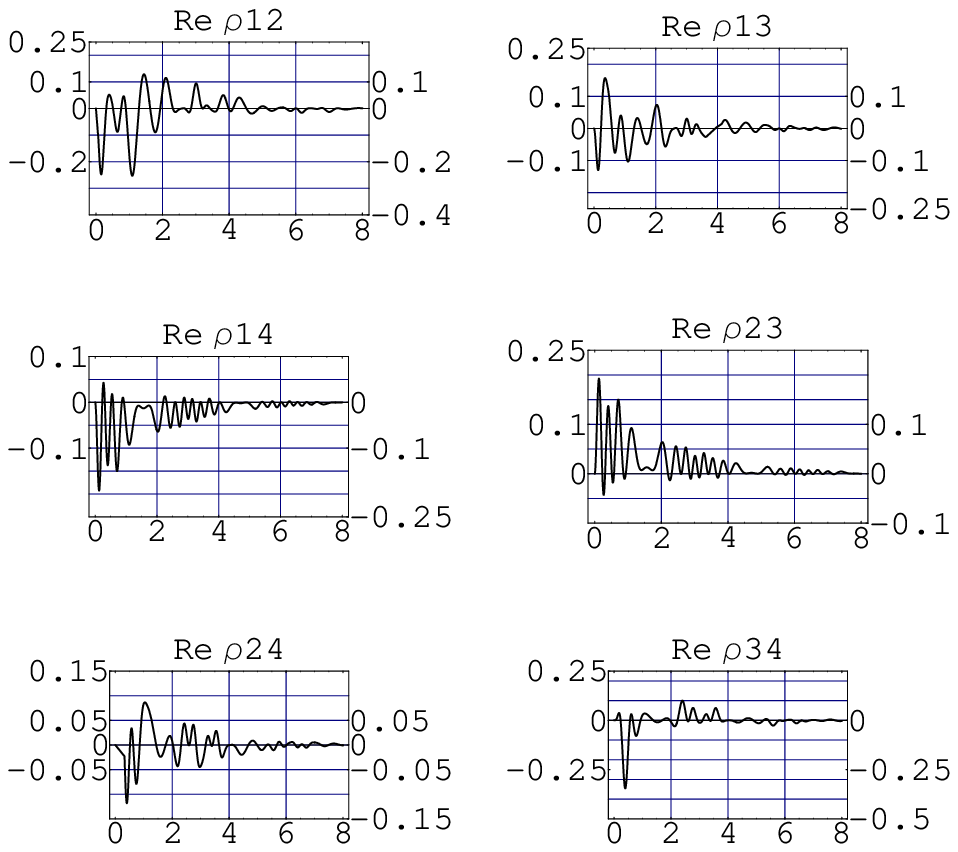}
\caption{As in Fig. 2 but with a damping constant $\Gamma = 0.5$ GHz. Note the differences from Fig. 2 as the oscillations damp out  asymptotically.}
\end{figure} 

\begin{figure}[h]
\includegraphics[width=3in]{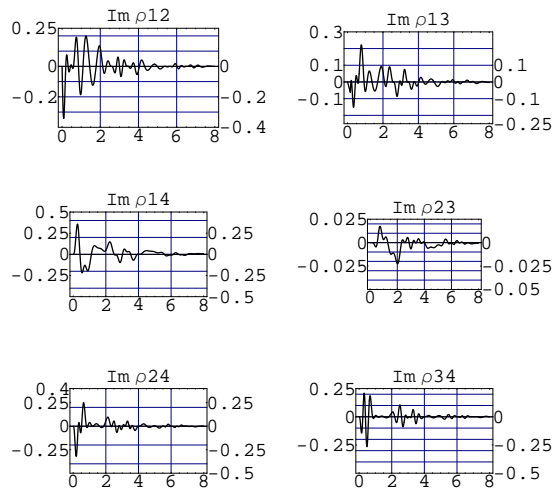}
\caption{As in Fig. 3 but with a damping constant $\Gamma = 0.5$ GHz. Note the differences from Fig. 3 as the oscillations damp out  asymptotically.}
\end{figure} 

Extension to decoherence and the master equation in Eq.~(\ref{eqn9}) requires handling the $15 \times 15$ problem in Eq.~(\ref{eqn10}) rather than $4 \times 4$ and is necessarily more involved numerically. However, as in earlier studies \cite{ref4,ref5}, a simplified model in which the sum in Eq.~(\ref{eqn9}) runs over all $L_k=O_k$ of Table 1 reduces to the same $4 \times 4$ calculation of unitary evolution above, with the only difference being to multiply $\eta (t)$ in Eq.~(\ref{eqn11}) by $\exp (-\Gamma t)$. Figs. 7-9 show the results of such a calculation for the same choice of parameters as in Figs. 1-3 but with the additional $\Gamma= 0.5$ GHz. The small $t$ dependence of the two sets of figures coincide but the oscillations are damped by the decoherence such that asymptotically, the density matrix evolves to that of a mixed state, namely diagonal components of 1/4 while all off-diagonal elements vanish. This evolution from a pure state to a completely mixed state is accompanied by a monotonic rise of the entropy from 0 to $\ln 4$, paralleling the similar results in \cite{ref4,ref5}.

This work has been supported by the U. S. Department of Energy and 
the National Science Foundation. One of us (ARPR) thanks the Tata 
Institute for Fundamental Research, Mumbai, India, for hospitality 
during the course of writing of this paper.


\begin{thebibliography}{}

\bibitem[*]{} Email: arau@phys.lsu.edu

\bibitem{ref1} F. Legare, Phys. Rev. A {\bf 68}, 063403 (2003); S. 
Jin, S. Gong, R. Li, and Z. Xu, Phys. Rev. A {\bf 69}, 023408 (2004).

\bibitem{ref2} V. Delgado and J. M. Gomez Llorente, Phys. Rev. A {\bf 
68}, 022503 (2003).

\bibitem{ref3} Yu. A. Pashkin, T. Yamamoto, O. Astafiev, Y. Nakamura, 
D. V. Averin, and J. S. Tsai, Nature {\bf 421}, 823 (2003); T. 
Yamamoto, Yu. A. Pashkin, O. Astafiev, Y. Nakamura, and J. S. Tsai, 
Nature {\bf 425}, 941 (2003).

\bibitem{ref4} A. R. P. Rau, Phys. Rev. Lett. {\bf 81}, 4785 (1998). 
See also, J. Wei and E. Norman, J. Math. Phys. {\bf 4}, 575 (1963); 
G. Dattoli, J. C. Gallardo, and A. Torre, Riv. Nuovo Cimento {\bf 
11}, No. 11, 1 (1988); G. Campolieti and B. C. Sanctuary, J. Chem. 
Phys. {\bf 91}, 2108 (1989); B. A. Shadwick and W. F. Buell, Phys. 
Rev. Lett. {\bf 79}, 5189 (1997).

\bibitem{ref5} A. R. P. Rau and R. A. Wendell, Phys. Rev. Lett. {\bf 
89}, 220405 (2002); A. R. P. Rau and Weichang Zhao, Phys. Rev. A {\bf 
68}, 052102 (2003).

\bibitem{ref6} S. Wolfram, \textit{Mathematica: A System for Doing 
Mathematics by Computer} (Addison-Wesley, Redwood City, CA, 1988).

\bibitem{ref7} A. W. Joshi, \textit{Elements of group theory for 
physicists} (John Wiley, New York, 1982); W. Greiner and B. M\"{u}eller, \textit{Quantum 
Mechanics: Symmetries} (Springer-Verlag, Berlin, 1989).

\bibitem{ref8} A. R. P. Rau, Phys. Rev. A {\bf 61}, 032301 (2000).

\bibitem{ref9} J. Zhang, J. Vala, S. Sastry, and K. B. Whaley, Phys. 
Rev. A {\bf 67}, 042313 (2003).

\bibitem{ref10} C. H. Bennett, D. P. DiVincenzo, J. Smolin, and W. K. Wootters, Phys. Rev. A {\bf 54}, 3824 (1996); Scott Hill and W. K. Wootters, Phys. Rev. Lett. {\bf 78}, 5022 (1997).

\bibitem{ref11} A formal description of this procedure is that for each of the 15 $O_i$, the set of $O_j$ obtained by the ST transformation, $O_j O_i O_j ^{\dagger}=O_i$, provide such a $su(2) \oplus su(2) \oplus u(1)$ sub-algebra. Similarly, the set satisfying $O_j O_i O_j ^{T}=-O_i$ provide the $su(2) \oplus su(2)$ or $so(5)$ sub-algebras, respectively, for $O_i$ symmetric or anti-symmetric.

\bibitem{ref12} See, for instance, D. F. Walls and G. Milburn, 
\textit{Quantum Optics} (Springer-Verlag, Berlin, 1994); M. O. Scully 
and M. S. Zubairy, \textit{Quantum Optics} (Cambridge University 
Press, Cambridge, U. K., 1996); W. P. Schleich, \textit{Quantum 
Optics in Phase Space} (Wiley-VCH, Berlin, 2001).

\bibitem{ref13} F. T. Hioe and J. H. Eberly, Phys. Rev. Lett. {\bf 47}, 838 (1981) and Phys. Rev. A. {\bf 25}, 2168 (1982).

\bibitem{ref14} R. G. Unanyan and M. Fleischhauer, Phys. Rev. A {\bf 69}, 050302(R) (2004).

\end{thebibliography}
\end{document}